\documentclass[sigconf]{acmart}
\AtBeginDocument{%
  }

\setcopyright{acmlicensed}
\copyrightyear{2025} 
\acmYear{2025} 
\acmConference[WWW '25]{Proceedings of the ACM Web Conference 2025}{April
	28-May 2, 2025}{Sydney, NSW, Australia}
\acmBooktitle{Proceedings of the ACM Web Conference 2025 (WWW '25), April
	28-May 2, 2025, Sydney, NSW, Australia}
\acmDOI{XXXXXXX.XXXXXXX}
\acmISBN{978-1-4503-XXXX-X/2018/06}

\usepackage{common}
\usepackage{mymacro}
\usepackage[hide]{notes}
\allowdisplaybreaks

\newif\ifsupp %
\supptrue %

\ifsupp\else  \fi %

\ifsupp
\usepackage[createShortEnv, conf={end, restate, text link=}]{proof-at-the-end}
\else
\usepackage[createShortEnv, conf={end, restate, text link=See proof in Appendix~A~[2].}]{proof-at-the-end}
\fi

\newcommand{\ilie}[1]{\textcolor{magenta}{}\xspace}
\newcommand{\aris}[1]{\textcolor{orange}{}\xspace}
\newcommand{\guangyi}[1]{\textcolor{blue}{}\xspace}

\newtheorem{problem}{Problem}

\newtheorem{example}{Example}

\crefname{algocf}{Algorithm}{Algorithms}

\begin{document}

\title[Efficient and Practical Approximation Algorithms for Advertising in Content Feeds]%
{Efficient and Practical Approximation Algorithms for\\Advertising in Content Feeds}

\author{Guangyi Zhang}
\affiliation{%
  \institution{Shenzhen Technology University}
  \city{Shenzhen}
  \country{China}
}
\email{zhangguangyi@sztu.edu.cn}
\orcid{0000-0002-1252-7489}

\author{Ilie Sarpe}
\affiliation{%
  \institution{KTH Royal Institute of
  	Technology}
  \city{Stockholm}
  \country{Sweden}}
\email{ilsarpe@kth.se}
\orcid{0009-0007-5894-0774}

\author{Aristides Gionis}
\affiliation{%
  \institution{KTH Royal Institute of Technology}
  \city{Stockholm}
  \country{Sweden}
}
\email{argioni@kth.se}
\orcid{0000-0002-5211-112X}

\begin{abstract}
Content feeds provided by platforms such as X (formerly Twitter) and TikTok are consumed by users on a daily basis.
In this paper, we revisit the native advertising problem in content feeds, initiated by Ieong et al.
Given a sequence of organic items (e.g., videos or posts) relevant to a user's interests or to an information search,
the goal is to place ads within the organic content
so as to maximize a reward function (e.g., number of clicks), 
while accounting for two considerations:
(1) an ad can only be inserted after a relevant content item; 
(2) the users' attention decays after consuming content or ads.
These considerations provide a natural model for capturing both the advertisement effectiveness and the user experience.
In this paper, we design fast and practical 2-approximation greedy algorithms for the associated optimization problem,
improving over the best-known practical algorithm that only achieves an approximation factor of~4. 
Our algorithms exploit a counter-intuitive observation, namely, 
while top items are seemingly more important due to the decaying attention of the user, 
taking good care of the bottom items is
key for obtaining improved approximation guarantees.
We then provide the first comprehensive empirical evaluation on the problem, 
showing the strong empirical performance of our~methods. 
\end{abstract}



\maketitle

\section{Introduction}\label{sec:intro}

A significant share of the current web traffic originates from user-generated content platforms,
such as X (formerly Twitter), Facebook, and TikTok~\cite{Alhabash2017Socials}.
These platforms primarily engage users through their \emph{content feeds}, 
which display a continuous stream of organic content items, 
such as social updates or videos,
arranged in a carefully crafted order and formatted for infinite scrolling~\citep{Milano2020Recc}.
The main monetization strategy of major social-media platforms is to insert sponsored content in between the content items, 
such as promoted posts, content seeking higher user engagement, 
or pay-per-click ads. 
The sponsored content is often designed to provide a well-integrated look and non-intrusive user experience, 
which is also known as \emph{native} advertising~\citep{wojdynski2016native}.
Advertisers incur a charge every time users interact with sponsored content, and
native advertising has evolved into a huge business with a market of
about 100 billion USD~\citep{stats2,stats3},
accounting for nearly two thirds of total display ad spending in the US~\citep{stats4}.

\begin{figure}[t]
	\centering
	
	\subcaptionbox{ }{ %
		\tikzstyle{edge1} = [yafcolor1, thick, >=latex]
\tikzstyle{edge2} = [yafcolor2, thick, >=latex]
\tikzstyle{edge3} = [-{Latex[length=0.8mm,width=1.2mm]}]
\tikzstyle{capt} = [left, above, draw=none, black, sloped, font=\scriptsize]
\begin{tikzpicture}[
	every node/.style={inner sep=1pt},
	]
	
	\node (v1) {$v_1$};
	\node (s1) [below=2.5mm of v1] {$s_1$};
	\node (v2) [below=2.5mm of s1] {$v_2$};
	\node (s2) [below=2.5mm of v2] {$s_2$};
	\node (p) [right=1.5mm of s2] {$(p)$};
	
	\node (a2) [left=18mm of s2] {$\ad_2$};
		
	\draw[edge1] (a2) edge node[capt]{\normalsize 2} (s2);
	
	\draw[edge3] (v1) to [in=45, out=-45] (v2);
	\draw[edge3] (v2) to [in=45, out=-45] (s2);
\end{tikzpicture}
	}
	\hfill
	\subcaptionbox{ }{ %
		\tikzstyle{edge1} = [yafcolor1, thick, >=latex]
\tikzstyle{edge2} = [yafcolor2, thick, >=latex]
\tikzstyle{edge3} = [-{Latex[length=0.8mm,width=1.2mm]}]
\tikzstyle{capt} = [left, above, draw=none, black, sloped, font=\scriptsize]
\begin{tikzpicture}[
	every node/.style={inner sep=1pt},
	]
	
	\node (v1) {$v_1$};
	\node (s1) [below=2.5mm of v1] {$s_1$};
	\node (v2) [below=2.5mm of s1] {$v_2$};
	\node (s2) [below=2.5mm of v2] {$s_2$};
	\node (p) [right=1.5mm of s2] {$(p' < p)$};
	
	\node (a1) [left=18mm of s1] {$\ad_1$};
	\node (a2) [left=18mm of s2] {$\ad_2$};
		
	\draw[edge2] (a1) edge node[capt]{\normalsize1} (s1);
	\draw[edge1] (a2) edge node[capt]{\normalsize 2} (s2);
	
	\draw[edge3] (v1) to [in=45, out=-45] (s1);
	\draw[edge3] (s1) to [in=45, out=-45] (v2);
	\draw[edge3] (v2) to [in=45, out=-45] (s2);
\end{tikzpicture}
	}
	\caption{\label{fig:decay}An illustration of the expected reward being non-monotone 
	with respect to the 
	ad placement.
	Here $a,v$ and $s$ denote ads, videos, and slots respectively.
	In the first scenario (a) an ad $a_2$ with reward 2 is allocated to slot $s_2$ after video~$v_2$, 
	and a user sees the ad $a_2$ with probability~$p$.
	In the second scenario (b) an additional ad $a_1$ with reward 1 is allocated to slot $s_1$ after video $v_1$.  
	Due to decaying user attention, 
	in (b), %
	the user sees the ad~$\ad_2$ with a probability $p'<p$. 
	Thus, placing an additional ad may lead to a smaller expected reward.}
\end{figure}
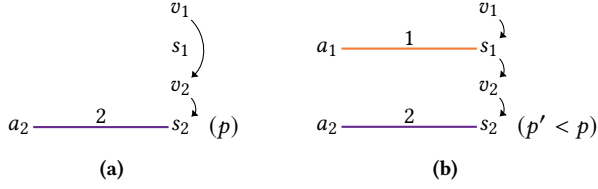

The placement of sponsored content within an infinite feed poses a unique allocation challenge
as it requires balancing two factors: 
(a) prioritizing advertisements at the top of the feed,
since users will eventually stop scrolling further %
their feed;
and (b) ensuring contextual coherence~\citep{yoon2023native},
to boost interaction rates. 
For instance, an airline advertisement is more attractive when displayed after a travel-related post rather than after a political one.
This setting is significantly different from traditional online advertising~\citep{mehta2013online,devanur2022online},
e.g., search advertising, 
where ads are sold through auctions for each opportunity, and
showing the winning ad is assumed to have no influence on a user session and future revenue.
In contrast, for native advertising in content feeds, 
showing an ad reduces the number of items a user will explore. %
Therefore, if no suitable advertisement fits a specific content, the optimal approach would be to forgo immediate revenue in favor of potential earnings later in the user session. %
For an illustration, consider \cref{example:decay} and \cref{fig:decay}.

\begin{example}\label{example:decay}
	As illustrated in \cref{fig:decay}, 
	assume that there is a \emph{slot} to which an ad can be allocated to, after every organic video.
	Consider two videos $v_1, v_2$ that are presented to a user in order.
	Suppose that an ad $\ad_2$ has been allocated to the slot after $v_2$.
	The crucial observation here is that 
	placing a new ad
	$\ad_1$ before $v_2$ 
	may lead to a loss in the total expected reward over the user session,
	as it reduces the probability that a user interacts with ad $\ad_2$.
\end{example}

\citet{ieong2014advertising} initiated a mathematical formulation for native advertising in content feeds,
denoted as the \streamads problem,
where in addition to %
given rewards for every feasible ad-item pair (e.g., collected through an ad auction),
users have %
decaying attention~\citep{craswell2008experimental}, and may quit browsing with a fixed probability after observing an item or an ad. 
Under such a model, the \streamads problem is to maximize the expected total reward over a user session, by suitably deciding a strategy to display ads.
\citet{ieong2014advertising} show that there exists a \PTAS (i.e., an algorithm that returns nearly optimal solutions) for the \streamads problem.
However, such an algorithm 
relies on solving expensive combinatorial problems, making it impractical.
To the best of our knowledge, the state-of-the-art practical algorithm only achieves a 4-approximation guarantee, %
that solves the problem by finding a suitable maximum weighted matching (\mwm) with %
cardinality constraints~\citep{ieong2014advertising}. 

In this paper, we develop practical and efficient 2-approximation greedy algorithms for the \streamads problem.
To deal with decaying attention,
our algorithms exploit a counter-intuitive structure of the problem, namely,
while top items are seemingly more important due to the decaying attention, 
finding a good position for the bottom items is key to obtaining improved approximation guarantees.
In addition, to carefully account for the constraints of \streamads, 
which require to allocate rewarding ads while considering the decaying attention of a user,
we devise a novel charging scheme based on a %
non-trivial decomposition of the \streamads's objective function. 
This result is then used to identify high-quality ad allocation strategies, and 
is leveraged in our proofs to obtain improved approximation~guarantees.

In addition, to the best of our knowledge, 
we provide the first comprehensive empirical study on the \streamads problem,
in which we verify the strong empirical performance of our novel algorithms. %

More specifically, our contributions are as follows.
\begin{itemize}
	\item We provide an exact greedy algorithm for a special case of the \streamads problem, 
	where each ad can be displayed more than once.
	\item We provide two 2-approximation greedy algorithms for the \streamads problem.
	The first algorithm uses a greedy criterion guided by the exact marginal gain in reward,
	and the second one leverages a lower bound of the marginal gain.
	The second algorithm is also particularly efficient in practice.
	\item We provide the first comprehensive empirical study on the \streamads problem, showing the high-quality ad allocations computed by our novel algorithms.
\end{itemize}

The rest of the paper is organized as follows.
We formally define the problem in \cref{sec:def}.
We characterize the structure of the problem in 
\cref{sec:case}.
We describe our novel algorithms and prove their approximation guarantees in \cref{sec:algs}.
Related work is discussed in \cref{sec:related} and
extensive experiments are in \cref{sec:exp}.
We conclude in \cref{sec:conclusion}. All the missing proofs are reported in~\cref{app:proofs}.

\section{Problem definition}\label{sec:def}
In this section, we first present the necessary preliminaries, and then formally define the problems 
studied in this paper.

\smallskip
\noindent
\emph{Preliminaries.}
A graph is \emph{bipartite} if its vertices can be partitioned into two disjoint parts, and
edges connect only vertices from different parts. 
Given an undirected graph, a \emph{matching} is a set of edges so that each vertex appears in at most one edge of the set.
For a weighted graph, a \emph{maximum-weight matching} (\mwm) is a matching in which the sum of its edge weights is maximized.

A set function $f: 2^\E \to \mathbb{R}$ %
assigns a value to every subset of a given set $\E$.
A set function $f$ is called \emph{monotonically non-decreasing} if $f(C) \le f(D)$,
for all $C \subseteq D \subseteq \E$.
Additionally, $f$ is called \emph{submodular} if $f(C + e) - f(C) \ge f(D + e) - f(D)$, 
for all $C \subseteq D \subseteq \E$ and element $e \in \E$.
Throughout this paper, we use the shorthands $C + e$ for $C \cup \{e\}$ and $C - e$ for $C \setminus \{e\}$.

An algorithm \ALG is an \emph{$\alpha$-approximation algorithm} for a maximization problem,  
if for any instance \instance of the problem, 
the solution $\ALG(\instance)$ returned by the algorithm
has an objective value that is no smaller than $1/\alpha$ 
times the value of the optimal solution, denoted with $\OPT(\instance)$~\citep{williamson2011design}.
That is, let $f$ be the objective function of the problem,
then it holds that $\alpha\, f(\ALG(\instance)) \ge f(\OPT(\instance))$, for all problem instances~\instance. 
A \emph{polynomial-time approximation scheme} (\PTAS) 
is an $(1+\varepsilon)$-approximation algorithm, for any given $\varepsilon > 0$,
with running time polynomial in the input size, but possibly exponential in~$1/\varepsilon$.

\smallskip
\noindent
\emph{Problem definition.}
We are given a sequence of \nV items (e.g., videos), 
and we assume that there is 
one available \emph{slot} for an ad placement after each item.
Suppose also that we are given \nads ads \ads. 
To improve the efficacy of the ads,
an ad $\ad_i$ can only be placed after a subset of relevant items $\slots_i \subseteq \V$.
A reward $\rw_{ij} \ge 0$ is then %
obtained if ad $\ad_i$ is shown to the user after the $j$-th item, with $j \in \slots_i$.
Throughout the paper, we fix $i$ (resp.~$j$) to be the index of an ad (resp.~a slot).

To model the decaying attention of the user, 
our model considers %
that a user decides to quit browsing (i.e., terminates their session) 
with probability $\q$ after observing every item or ad.
Our goal is to decide the allocation of ads to the available slots 
to maximize the expected reward over the specified model. 
We use the terms reward and revenue interchangeably.
For brevity, we may drop the adjective ``expected'' if it is clear from the context.
More formally, the ad-placement problem is defined as~follows.

\begin{problem}[\streamadsr]\label{problem:adsr}
	We are given a sequence of \nV items~\V with one available slot after each item, 
	a set of \nads ads $\ads = \{ \ad_i \}$ with associated slots $\{ \slots_i \}$,  
	rewards $\{ \rw_{ij} \}$ for $j \in \slots_i$, and a quitting probability  $\q\in[0,1)$.
	The goal is to find a mapping 
	$\M \subseteq \E := \bigcup_{i \in [\nads]} \left(\{ i \} \times \slots_i\right)$ 
	such that every slot can admit at most one ad, i.e.,
	$|\{ i : (i, j) \in \M \}| \le 1$ for all~$j$, 
	and $\M$ maximizes the expected reward
	\begin{align}
		f(\M) := \sum_{e=(i,j) \in \M} \rw_{e} (1-q)^{j + \nb(j)}, \label{eq:obj}
	\end{align}
	where $\nb(j)$ is the number of slots before slot $j$ containing an ad,
	i.e.,
	$\nb(j) = |\{ j' < j : (i, j') \in \M \text{ for some } i \}|$.
\end{problem}

The \streamadsr problem explicitly disallows consecutive ads, 
which helps to avoid ad fatigue
and viewer zapping~\citep{shi2023much}.
\streamadsr also benefits from state-of-the-art recommenders that can be used to obtain high-quality rankings for the content items, as it is a common practice to design  ad-allocation strategies
as a post-processing operation~\citep{yan2020ads,li2024deep}. 
In addition, state-of-the-art machine learning models can also be used to obtain high-quality predictions for the expected rewards $r_e$ from~\cref{eq:obj} over user sessions, e.g., from historical data. %

The \streamadsr problem allows an ad to be displayed multiple times.
However, there are scenarios where displaying an ad multiple times is undesirable.
To prevent such over-exposure of ads,
it is possible to preprocess the slots $\slots_i$ of each ad $\ad_i$
and set a limit on the number of slots $|\slots_i|$.
However, such an approach is limited and not always feasible.
To provide a rigorous model for such cases, 
we introduce the following problem variant.

\begin{problem}[\streamads]\label{problem:ads}
	Given the same input as in the \streamadsr\ problem,
	find a \emph{matching} $\M \subseteq \E := \bigcup_{i \in [\nads]} \left(\{ i \} \times \slots_i\right)$ 
	that maximizes the expected reward $f(\M)$ from~\cref{eq:obj}. %
\end{problem}

Note that the \streamads problem is significantly more general than the previous \streamadsr problem, 
as an ad can be 
displayed multiple times also in \streamads, by simply generating multiple copies of such an ad.
Besides, the \streamads problem also generalizes the classic maximum-weight matching problem (\mwm),
obtained from \streamads by setting the value $\q=0$.
\citet{ieong2014advertising} also prove that there is no \emph{online} algorithm with a constant competitive ratio for \streamads. 
Hence we focus on the offline settings. 

Finally, in \cref{sec:exp:ablation}
we also discuss how to adapt an algorithm for \streamads 
to %
enforce a limit on the total number of ads to be displayed,
which can be useful, for example, to avoid ad fatigue. %

\section{Problem structure and failed attempts}\label{sec:case}
The \streamads problem was introduced by \citet{ieong2014advertising},
who also devised a \PTAS algorithm.
However, their \PTAS relies on exhaustive enumeration of sub-sequences of slots, and flow computations, which is impractical. %
In this section, we study the structural properties of the \streamads problem
aiming to design a practical algorithm with provable quality~guarantees.

Our first step is to view %
the \streamads problem
as a task of optimizing a specific set function over a bipartite matching.
However, as shown in \cref{prop:nonsubm}, 
this specific set function is neither monotone nor submodular.
Therefore, the problem cannot be approximated 
by existing methods for submodular maximization~\citep{buchbinder2018submodular}.

We then present an example showing that two simple and intuitive heuristics may perform arbitrarily bad.
The first heuristic is a standard greedy strategy that prioritizes placing ads in the \emph{top slots}, 
i.e., the slots appearing at the beginning of the content feed.
The second heuristic is to address the problem leveraging the %
maximum-weight matching (\mwm) method. 
The failure of such approaches, and the problem instance that causes the
two heuristics to perform badly inspire the design of our novel algorithms. 
In the next section	(\cref{sec:algs}) we 
propose
a novel backwards greedy strategy 
that carefully accounts for the placement of ads in \emph{bottom slots}, i.e., the slots appearing at the \emph{end} of the content feed.

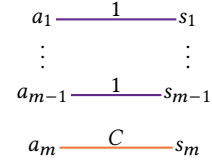
\begin{figure}[t]
	\centering
	\tikzstyle{edge1} = [yafcolor1, thick, >=latex]
\tikzstyle{edge2} = [yafcolor2, thick, >=latex]
\tikzstyle{capt} = [left, above, draw=none, black, sloped, font=\scriptsize]
\begin{tikzpicture}[
	every node/.style={inner sep=0.7pt},
	]
	
	\node (c2) {$\svdots$};
	\node (c1) [above=2mm of c2] {$\ad_1$};
	\node (c3) [below=2mm of c2] {$\ad_{\nV-1}$};
	\node (t2) [right=18mm of c2] {$\svdots$};
	\node (t1) [above=2mm of t2] {$s_1$};
	\node (t3) [below=2mm of t2] {$s_{\nV-1}$};
	
	\node (a) [below=4.5mm of c3] {$a_\nV$};
	\node (b) [below=4.5mm of t3] {$s_\nV$};
	
	\draw[edge1] (c1) edge node[capt]{\normalsize1} (t1);
	\draw[edge1] (c3) edge  node[capt]{\normalsize1} (t3);
	\draw[edge2] (a) edge node[capt]{\normalsize$C$} (b);
\end{tikzpicture}
	\caption{Representation of \cref{exmaple:top}, where %
		a natural online greedy algorithm and 
		maximum weighted matching (\mwm)
		perform poorly ($\ad$ and $s$ represent ads and slots).}\label{fig:badcase}
\end{figure}

\begin{propositionE}\label{prop:nonsubm}
	The expected-reward function $f: 2^\E \to \reals$ in \cref{eq:obj} for the \streamads problem is 
	neither monotone nor submodular.
\end{propositionE}
\begin{proofE}
	For simplicity, we consider a special case where,
	for each ad $\ad_i$, the rewards $r_{ij}$ are identical, i.e., $r_{ij} = r_i$, 
	for all associated slots $j \in \slots_i$.
	We first show that the expected reward is non-monotone. 
	It is easy to see that assigning ads sequentially by the order of the slots increases the expected reward.
	However, assigning a new ad with a zero reward to an earlier slot decreases the expected reward, as it reduces the probability of subsequent ads of being seen.
	
	We continue to show that the expected-reward function is non-submodular. 
	For any feasible subset $C \subseteq D \subseteq \E$, 
	the marginal gain $g((i,j) \mid C) = f(C+(i,j)) - f(C)$ 
	of adding an edge $(i, j)$ into a set of edges~$C$ is
	\begin{align*}
		g((i, j) \mid C) 
		= \rw_i (1-\q)^{j + \nb(j)} - \q \sum_{(i',j') \in C: j' > j} \rw_{i'} (1-\q)^{j' + \nb(j')}.
	\end{align*}
	Compared with $g((i ,j) \mid D)$, 
	the first term is clearly non-increasing,
	but the second term may increase.
	For example, we have $g((i ,j) \mid C) < g((i ,j) \mid D)$ 
	by letting $D \setminus C$ be ads with zero rewards placed after slot $j$ \emph{and} before other subsequent items.
	On the other hand, 
	we also have $g((i ,j) \mid C) \ge g((i ,j) \mid D)$ 
	when slot $j$ is ranked after every occupied slot in $D$.
\end{proofE}

Due to the exponentially-decaying attention in the model, 
a reasonable strategy is to prioritize the top slots.
Thus, a logical choice is to employ a greedy algorithm that processes slots in a sequentially increasing order 
and repeatedly matches the ad with the highest reward to the processed slot.
However, as we show below, 
such a greedy algorithm has an unbounded approximation ratio, even for the easier \streamadsr problem.

\begin{example}[Being myopic in top slots]\label{exmaple:top}
	See \cref{fig:badcase} for an illustration.
	For each slot $j = 1, \ldots, \nV-1$, we create a dedicated ad $\ad_j$ with reward~1.
	For the final slot $j=\nV$, we create an ad $\ad_j$ with a large reward~$C$.
	The greedy algorithm assigns each ad in its corresponding slot, and it results 
	in a total expected reward of 
	\begin{equation*}
		\sum_{j=1}^{\nV-1} (1-\q)^{2j-1} +  (1-\q)^{2\nV-1} C
		 \approx \tfrac{(1-\q)}{1-(1-\q)^2} + (1-\q)^{2\nV-1} C.
	\end{equation*}
	On the other hand, assigning only the last ad gives reward $(1-\q)^{\nV-1} C$.
	For certain values of the parameters 
	the approximation ratio can be arbitrarily bad.
	For example, when $\q=1/2$ and $C=2^{2\nV-1}$,
	the approximation ratio is about $2^\nV / 2$. 
\end{example}

The instance in \cref{exmaple:top} is also hard for another intuitive algorithm based on maximum-weight matching (\mwm).
This algorithm finds a \mwm for the bipartite graph between ads and slots with appropriately-defined edge weights.
That is, every edge $(i,j)$ connecting ad $\ad_i$ and slot $j$ has a position-biased weight of $\rw_{ij} (1-q)^{j}$.
Unfortunately, the \mwm algorithm fails to capture the decaying-attention effect of the model.
It is easy to see that, on the instance from \cref{exmaple:top}, the \mwm algorithm selects all available edges, like the afore\-mentioned greedy~algorithm.

By a careful inspection of the bad instance in \cref{exmaple:top}, 
it is clear that to obtain solutions with high expected reward, 
we cannot only focus on the top slots, 
or
ignore the decaying-attention effect of the model.
However, it is difficult to take care of both ends of the slot sequence. %
We show in the next section, that both issues can be handled properly by %
first considering 
bottom~slots, through our novel algorithms.

\section{Algorithms}\label{sec:algs}
In this section, we introduce a novel backwards-greedy algorithm
(\cref{alg:greedy-back}, denoted as \alggback) that carefully handles the bottom slots
for ad placement.
The backwards-greedy approach addresses %
the decaying-attention in the model, 
by iteratively considering
sub-problems over suffixes (of the form $j,\dots,\nV$, for \emph{decreasing} $j$) of the slots. %
We show in \cref{thm:greedy-streamadsr} that the backwards-greedy algorithm, perhaps surprisingly, 
finds an optimal solution for the \streamadsr~problem.

On the other hand, 
it is not straightforward to
analyze the \alggback algorithm for the more challenging \streamads problem 
due to the interplay between the decaying-attention effect and the additional matching constraint. %
To address this issue, we prove a novel decomposition of the expected reward over a matching, 
which we use to obtain a \emph{non-oblivious} backwards-greedy 2-approximation algorithm
(\alggbackproxy in \cref{alg:greedy-backproxy}) for the \streamads problem, 
running much faster than \alggback.
More specifically, 
the \alggbackproxy algorithm adopts a greedy criterion that deviates from the standard marginal-gain greedy criterion (with respect to the underlying objective value). %
Finally, by leveraging the structural lemmas for the \alggbackproxy algorithm, 
we provide an analysis for the \alggback algorithm. %
We conclude by also presenting other practical algorithms that can be used to solve \streamads.

Before presenting our novel algorithms,
we
introduce a sub-problem of \streamads, 
which we refer to as \streamadsj, for a fixed integer~$j\in[\nV]$.
In the \streamadsj sub-problem,
the first $j$ items and slots are not considered, i.e., we only consider slots $j+1,\dots,m$.
The resulting objective function for \streamadsj is,
\begin{align}
	f_j(\M) := \sum_{e=(i,j') \in \M: j' > j} \rw_{e} (1-q)^{j'-j + \nb_j(j')}, \label{eq:obj-j}
\end{align}
where $\nb_j(j')$ is the number of slots after slot $j$ and before slot $j'$ containing an ad,
i.e.,
$\nb_j(j')= |\{ j < k < j' : (i, k) \in \M \text{ for some } i \}|$.
In~particular, 
$f_0 = f$, while 
$f_{\nV}(\cdot) = 0$. %

\subsection{Solving \streamadsr optimally}\label{sec:algs:streamadsr}
\begin{algorithm}[t] %
	\SetKwComment{tcp}{$\triangleright$\ }{}%
	\SetCommentSty{small}
	$\M \gets \emptyset$; $\ads_j \gets \{ i : j \in \slots_i \}$ for all $j\in[m]$\;
	\For{slot $j = \nV, \ldots, 1$ (in a reverse order)}{
		\For{$i \in \ads_j$}{
			$\M_i \gets \M + (i, j)$\;
			\If{$M_i$ is not a valid matching for \streamads }{
				$\M_i \gets (\M \setminus \{ (i, j'): j' \in [\nV] \}) + (i, j)$\label{alg:bwd:re-assign}\;
			}
			$g_i \gets f_{j-1}(\M_i) / (1-\q) - f_j(\M)$\label{eq:g}\;
		}
		$i^* \gets \arg\max_{i \in \ads_j} \{g_i\}$\;
		\lIf{$g_{i^*} > 0$}{$\M \gets \M_{i^*}$}
	}
	\Return \M\;
	\caption{Backwards greedy (\alggback)}
	\label{alg:greedy-back}
\end{algorithm}

Our  backwards-greedy algorithm (\alggback) 
for both the \streamadsr and \streamads problems
is illustrated in \cref{alg:greedy-back}.
The \alggback algorithm returns an optimal solution for the \streamadsr problem, 
as we prove in \cref{thm:greedy-streamadsr}.

The \alggback algorithm processes the slots in a reverse order, starting from the final slot.
At each slot, \alggback tries to (re-)assign an ad
by finding the ad that maximizes the \emph{marginal gain} for the revenue (defined in \cref{eq:g}).
The algorithm performs a (re-)assignment if it results in a positive marginal gain (i.e., increasing the objective function).
A matching (or a mapping for \streamadsr) is then returned after processing all slots.

\begin{theoremE}\label{thm:greedy-streamadsr}
\cref{alg:greedy-back} solves the \streamadsr problem optimally.
\end{theoremE}
\begin{proofE}
	The proof is similar to the one by \citet{ieong2014advertising} 
	for finely targeted ads, i.e., $|\slots_i| = 1$, for all ads $\ad_i$.
	The key is to notice that by processing slots backwards, 
	a decision 
	at slot $j$
	cannot affect any slot 
	that has not yet been processed, i.e., slots in positions $j'=1,\dots,j-1$.
	That is, the user attention for a slot $j'$ does not depend on ads placed later (in slots $j,\dots,\nV$);
	additionally, 
	every ad can be re-used as there is no matching constraint.
	Thus, 
	solving optimally the sequence of sub-problems on slots $j,\dots,\nV$ 
	with decreasing $j=\nV,\dots,1$,
	yields an optimal solution to \streamadsr.
	
	The sub-problem for the final slot (i.e., $j=\nV$) is trivial, 
	and \alggback assigns to it the ad with the highest expected reward, 
	if available.
	Moving backwards to the next slot $j$, 
	\alggback assigns an ad with the highest reward to the slot $j$ only if it improves the total reward,
	that clearly
	results in an optimal assignment for this new sub-problem. %
	The proof immediately follows by the above invariant over the backward processing of the slots. 
\end{proofE}

The time complexity for the \alggback algorithm is 
$\bigO(|\E|)$ for \streamadsr, and
$\bigO(|\E| \beta) = \bigO(|\E| \min \{ \nV, \nads \})$ for \streamads,
where $\beta = \bigO(|\M|)$ is the time used to compute~$f_j(\M)$ for $j\in [\nV]$.

\subsection{Non-oblivious greedy for \streamads}\label{sec:algs:streamads}
\begin{algorithm}[t] %
	\SetKwComment{tcp}{$\triangleright$\ }{}%
	\SetCommentSty{small}
	$\M \gets \emptyset$; $\tau_i \gets 0$ for all $i$;
	$\ads_j \gets \{ i: j \in \slots_i \}$ for all $j$\;
	\For{slot $j = \nV, \ldots, 1$ (in a reverse order)}{
		\lIf{it exists $j' \text{ s.t.\ } (i, j') \in \M$}{$\sigma(i) \gets j'$}
		\lElse{$\sigma(i) \gets j$}
		$i^* \gets \arg\max_{i \in \ads_j} \{\rw_{ij} - \tau_i (1-\q)^{\sigma(i)-j}\}$\label{step:greedy-proxy}\;
		$g_{\text{LB}} \gets \rw_{i^*j}  -\q f_j(\M) - \tau_{i^*} (1-\q)^{\sigma(i^*)-j}$\label{step:LB}\;
		\If{$g_{\text{LB}} > 0$}{
			$\M \gets (\M \setminus \{ (i^*, j'): j' \in [\nV] \}) + (i^*, j)$\tcp*{(re-)assign $a_{i^*}$}
			$\tau_{i^*} \gets \rw_{i^* j} - \q \, f_j(\M)$\;
			\If{$\ad_{i^*}$ is re-assigned}{
				$\tau_{i} \gets \rw_{ij'} - \q \, f_{j'}(\M)$ for every $(i,j') \in \M$\;
			}
		}
	}
	\Return \M\;
	\caption{Non-oblivious backwards greedy (\alggbackproxy)}
	\label{alg:greedy-backproxy}
\end{algorithm}

The \streamads problem is more challenging due to the matching constraint.
A first idea to address such a problem would be to leverage the~\alggback algorithm, 
and decompose the reward of a matching 
into a sum of marginal gains, %
one term for each slot.
Then, 
to provide approximation guarantees, we need to 
connect such marginal rewards to those of an optimal solution for \streamads. %
However, such analysis quickly becomes challenging, 
as a single re-assignment (in \cref{alg:bwd:re-assign}) may affect the marginal gain over multiple slots due to the decaying-attention~effect.

To avoid such issues, we relate the total revenue to a lower bound of the marginal gains in the above decomposition,
that we use to develop a novel greedy algorithm.
This results in a 2-approximation 
non-oblivious backwards-greedy algorithm 
(\alggbackproxy in \cref{alg:greedy-backproxy}) 
for the \streamads problem, note that
this approximation ratio is tight for any greedy algorithm.
The \alggbackproxy algorithm is called ``\emph{non-oblivious}''~\citep{khanna1998syntactic} 
since
it does not select the next ad
with respect to the objective function $f$ of \cref{eq:obj}.

\smallskip
\noindent
\emph{The \alggbackproxy algorithm.}
The \alggbackproxy algorithm is introduced in \cref{alg:greedy-backproxy}.
Similar to the \alggback algorithm,
it processes the slots in a reverse order, starting from the final slot.
The key difference is that,
at every slot $j=\nV,\dots,1$, it seeks to (re-)assign an ad that maximizes a \emph{lower bound} of the marginal gain, 
which is
\begin{align}
	\arg\max_{i \in \ads_j} \; \left\{\rw_{ij} - \q \, f_j(\M) - \tau_i (1-\q)^{\sigma(i)-j}\right\}, \label{eq:greedy-proxy}
\end{align}
where 
$\tau_i$ is defined below,
$\ads_j = \{ i: j \in \slots_i \}$, and
$\sigma(i)=j$ %
if $\ad_i$ is new to the matching \M, 
otherwise $\sigma(i)$ corresponds to the slot previously selected for $\ad_i$.
We prove shortly (in \cref{lemma:LB}) that \cref{eq:greedy-proxy} is a lower bound 
to the marginal reward obtained by assigning an ad at slot~$j$. 

The term $\tau_i$ represents an \emph{estimate} of the total prior reward provided by ad $\ad_i$.
At the beginning, $\tau_i$ is initialized to be 0 for all $i$.
Every time an ad $\ad_i$ is (re-)assigned to the $j$-th slot,
we update its value according to the following rule:
\begin{align}
	\tau_i = \rw_{ij} - \q \, f_j(\M). \label{eq:tau}
\end{align}
It is easy to see that, 
the first time an ad $\ad_i$ is assigned to the slot $j$,
$\tau_i$ represents its actual marginal gain.
However, afterwards, if the ad $\ad_i$ is re-assigned to a different slot $j'<j$,
$\tau_i$ deviates from its marginal gain 
as it does not consider the variation over
$f_{j'}(\M)$, 
caused by the withdrawal of $\ad_i$ from slot $j$.
During the execution of \alggbackproxy, 
it is important to maintain each $\tau_i, i\in[n]$ 
up-to-update when re-assignments occur.
We write $\tau_{j} = \rw_{e_{j}} - \q \, f_j(\M)$ 
when it is more convenient to use the \emph{slot} index $j$,
where $e_j$ denotes an edge that is assigned to the slot $j$.

The \alggbackproxy algorithm preforms an ad (re-)assignment if it results in a positive lower bound as from \cref{eq:greedy-proxy}.
A matching $\M$ is returned after processing all slots. 
We show in \cref{thm:greedy-streamads} that it holds $2 f(\M) \ge f(\M^*)$, where $\M^*$ is the matching achieving the optimal solution for $f$, i.e., \alggbackproxy is a 2-approximation algorithm.

\smallskip
\noindent
\emph{Decomposition.}
We now introduce the novel decomposition of the reward of a matching \M.
Let $R_j := f_j(\M)$ be the reward of a solution \M for the \streamadsj sub-problem (from \cref{eq:obj-j}).
We have
\begin{align}
	R_j 
	&= (1-\q) \left( R_{j+1} + \indicator[e_{j+1} \in \M] (\rw_{e_{j+1}} - \q R_{j+1}) \right) \label{eq:recursion}\\
	&= \sum_{j'=j+1}^{\nV}  (1-\q)^{j'-j} \indicator[e_{j'} \in \M] (\rw_{e_{j'}} - \q R_{j'}) \label{eq:decomp} \\
	&= \sum_{e=(i,j') \in \M: j' > j}  (1-\q)^{j'-j} (\rw_{e} - \q R_{j'}) \nonumber
	,
\end{align}
where $\indicator[e_j \in \M]$ is a 0--1 indicator function taking value 1 if the edge $e_j$, incident to slot $j$, is in the matching \M.
The first equality (\cref{eq:recursion}) expresses $R_j$ as a sum of $R_{j+1}$, and the marginal gain obtained by allocating slot $j+1$ with edge $e_{j+1}$.
The second equality (\cref{eq:decomp}) recursively expands the term $R_{j+1}$, while groups the other terms into a summation. %
The last equality follows 
a simple double-counting argument. %
In summary, $R_j$ is a cumulative sum of marginal gains, computed backwards, of edges in \M,
when there are \emph{no} re-assignments.
Notice the similarity between the components in the decomposition in \cref{eq:decomp}
and the values $\tau_i$ in \cref{eq:tau} (recall that $R_j = f_j(M)$).

We next characterize the behaviour of $R_j$ when a re-assignments occurs in the backwards-greedy algorithm,
and connect such results to the greedy criterion in \cref{eq:greedy-proxy}.

\begin{lemmaE}\label{lemma:R}
	During the execution of the main loop of~\cref{alg:greedy-backproxy},
	for any fixed $j \in [\nV]$,
	the value $R_j$ is non-increasing 
	since the completion of the sub-problem \streamadsj.
	\end{lemmaE}
\begin{proofE}
	At each iteration, $R_j$ remains unchanged if no re-assignment occurs.
	Hence, consider when an ad $\ad_i$ 
	is re-assigned from slot $\tilde{j}$ to slot $\tilde{j}'$, and
	let $\tilde{R}_j$ be the revenue after such a re-assignment.
	First note that our statement does not regard $R_{\tilde{j}'}$, 
	because the sub-problem \streamads-$\tilde{j}'$ is completed after the re-assignment.
	Clearly, $\tilde{R}_{j} = R_{j}$ for any $j \ge \tilde{j}$.
	Now let $j < \tilde{j}$. 
	We prove by induction that $\tilde{R}_{j} \le R_{j}$.
	Recall that $\tau_{j} = \rw_{e_{j}} - \q R_{j}$,
	and by design of the \alggbackproxy algorithm it holds $\tau_{j} > 0$. 
	
	First, as a base case, when $j = \tilde{j} - 1$, we have
	\[
	\tilde{R}_{j}  = (1-\q) R_{\tilde{j}} 
	\le (1-\q) (R_{\tilde{j}} + \tau_{\tilde{j}})
	= R_{j}.
	\]
	In the inductive step, for $j< \tilde{j} -1 $, we have
	\begin{align*}
		\tilde{R}_{j}  &= (1-\q) (\tilde{R}_{j+1} + \indicator[e_{j+1} \in \M - (i,\tilde{j})] \tilde{\tau}_{j+1}) \\
		&\le (1-\q) (R_{j+1} + \indicator[e_{j+1} \in \M] \tau_{j+1})
		= R_{j},
	\end{align*}
	where $\tilde{\tau}_{j} = \rw_{e_{j}} - \q \tilde{R}_{j}$.
	The inequality follows since $\tilde{R}_{j+1} \le R_{j+1}$ holds regardless of $e_{j+1}$ being in $\M$ or not.
	This completes the proof.
\end{proofE}

\smallskip
\noindent
\emph{Approximation guarantees.}
Next, we explain the novel lower bound presented in \cref{eq:greedy-proxy}.
When re-assigning an ad, the exact marginal gain in reward 
heavily depends on the allocation of all other slots already allocated,
due to the decaying attention,
making the analysis particularly challenging.
Therefore, instead of considering the actual marginal reward, %
\alggbackproxy seeks a greedy choice that
maximizes the non-oblivious lower bound, 
which simplifies our analysis.
We first prove that \cref{eq:greedy-proxy} (evaluated by~\alggbackproxy in \cref{step:LB}) 
is a lower bound to the actual marginal reward,
provided that every $\tau_i$ in \cref{eq:tau} is maintained up-to-update.

\begin{lemmaE}\label{lemma:LB}
	Denote by $g$ the marginal gain in reward of re-assigning ad $\ad_i$ from slot $\tilde{j}$ to slot $j$ with $\tilde{j} > j$.
	Then,
	\[
		g \ge \rw_{ij} -\q R_j - \tau_{i} (1-\q)^{\tilde{j}-j}.
	\]
\end{lemmaE}
\begin{proofE}[no link to proof]
	The marginal gain $g$ of re-assigning ad $\ad_i$ from slot $\tilde{j}$ to slot $j$ is a sum of two terms.
	The first term is the loss of removing edge $\tilde{e} = (i,\tilde{j})$, and
	the second term is the marginal reward of adding the new edge $(i,j)$.
	By \cref{eq:decomp}, we have that 
	\begin{align*}
		R_j - \tilde{R}_j 
		&= \sum_{j'=j+1}^{\nV}  (1-\q)^{j'-j} \left( 
			\indicator[e_{j'} \in \M] \tau_{j'}
			- \indicator[e_{j'} \in \M - \tilde{e}] \tilde{\tau}_{j'}
		\right) \\
		&= \tau_{i} (1-\q)^{\tilde{j}-j} +
		\sum_{j'=j+1}^{\tilde{j}-1}  (1-\q)^{j'-j} \left( 
		\indicator[e_{j'} \in \M] (\tau_{j'} - \tilde{\tau}_{j'})
		\right) \\
		&\le \tau_{i} (1-\q)^{\tilde{j}-j} 
		,
	\end{align*}
	where $\tilde{R}_j$ is the reward after the removal, and 
	$\tilde{\tau}_{j} = \rw_{e_{j}} - \q \tilde{R}_{j}$.
	The last two steps follow from \cref{lemma:R}.
	The claim follows,
	\begin{align*}
		g &= \tilde{R}_j - R_j + \rw_{ij} -\q \tilde{R}_j 
		= (1-\q) (\tilde{R}_j - R_j) + \rw_{ij} -\q R_j \\
		&\ge \rw_{ij} -\q R_j - \tau_{i} (1-\q)^{\tilde{j}-j+1}
		\ge \rw_{ij} -\q R_j - \tau_{i} (1-\q)^{\tilde{j}-j}
	\end{align*}
\end{proofE}

Finally, we are ready to show the approximation ratio for \cref{alg:greedy-backproxy}.

\begin{theoremE}\label{thm:greedy-streamads}
	\cref{alg:greedy-backproxy} returns a 2-approximation for the \streamads problem.
\end{theoremE}
\begin{proofE}[no link to proof]
	We prove the claim by induction on slots $j\in[\nV]$ following the same backward ordering (i.e., $j=\nV,\dots, 1,0$) adopted by \cref{alg:greedy-backproxy}. 
	Let $\ALG_j$ be the 
	solution of \cref{alg:greedy-backproxy} before
	performing the $j$-th iteration
	(i.e., having only processed the slots in positions $\nV,\dots, j+1$)\footnote{for $j=\nV$ there are no such processed slots, while if $j=0$ then $\ALG_j$ corresponds to the output of~\cref{alg:greedy-backproxy}.}, and 
	$\OPT_j$ be the optimal solution to \streamads (i.e., \OPT) ignoring the first $j$ slots.
	Let their objective values for the sub-problem \streamadsj be $R_j := f_j(\ALG_j)$ and $R_j^* := f_j(\OPT_j)$, respectively.
	And also let the marginal revenue in $R$ be $g_j = R_{j-1} / (1-\q) - R_j$ at the $j$-th slot, and similarly in $R^*$, $g^*_j = R^*_{j-1} / (1-\q) - R_j^*$.
	We then write $\Gamma_i := \tau_i (1-\q)^{\sigma(i)-j}$,
	for each ad $a_i$ matched in $\ALG_j$.
	
	Let $\tilde{j}$ be smallest $j$ such that it holds
	$R_{\tilde{j}} \ge R_{\tilde{j}}^*$.
	Note that $\tilde{j}$ exists, as $R_\nV = R_\nV^* = 0$.
	If $\tilde{j} = 0$, 
	the statement trivially follows.
	Otherwise, we assume the following hypothesis:
	for every $j < \tilde{j}$,
	we can charge the marginal revenue 
	$g^*_j$ of $\OPT_j$ to both $g_j$ and $\{ \Gamma_i \}$ in $\ALG_j$,
	while maintaining the invariant that
	every $\Gamma_i$ (corresponding to ad $\ad_i$) in $\ALG_j$ is used at most once among all iterations.
	This immediately implies
	\begin{align*}
		2 f_j(\ALG_j) &=  \sum_{j' > j} g_{j'} (1-\q)^{j'-j} + \sum_{e=(i,j') \in \ALG_j}  \Gamma_{i} \\
		&\ge \sum_{j' > j} g^*_{j'} (1-\q)^{j'-j}
		= f_j(\OPT_j)
	\end{align*}
	by the decomposition in \cref{eq:decomp}.
	
	For $j = \tilde{j}$, 
	since $R_j \ge R_j^*$,
	it is sufficient to consider only the marginal gains
	$\{ g_j \}$, as it holds $R_j \ge R_j^*$.
	Now, for the next smaller $j$ in an inductive step, we have the following cases.
	
	\textbf{Case 1}. $\OPT_{j-1} = \OPT_{j}$, that is, \OPT does not include any new ad for its $j$-th slot.
	If our \ALG also does not select any item for the $j$-th slot, then the inductive step clearly holds.
	
	Otherwise, notice that~\cref{alg:greedy-backproxy} (re-)assigns an ad only if $g_{LB} > 0$ by \cref{lemma:LB}.
	Hence, the overall revenue (i.e., $R_{j-1}/(1-\q)$) only increases, 
	and therefore our hypothesis holds also for this case.
	
	\textbf{Case 2}. $\OPT_{j-1} = \OPT_{j} + e^*$, where $e^* = (i^*, j)$, that is the optimal solution assigns ad $i^*$ to the $j$-th slot.
	
	\textbf{Case 2.1}. If our \ALG (re-)assigns ad $i$ to slot $j$, i.e.,  matching the edge $e=(i, j)$, then by the greedy criterion (\cref{eq:greedy-proxy}), we have
	\[
	\rw_e - \Gamma_i \ge \rw_{e^*} - \Gamma_{i^*} .
	\]
	Therefore, we can use both $\Gamma_{i^*}$ and $g_j$ to charge for $\rw_{e^*}$. 
	That is,
	\begin{align*}
		g_j + \Gamma_{i^*} 
		&\ge \rw_e - \Gamma_{i} - q R_j + \Gamma_{i^*} %
		\ge \rw_{e^*} - q R_j^* = g^*_j,
	\end{align*}
	where the first inequality follows by \cref{lemma:LB}, and
	the second follows by the greedy rule and the fact that $R_j < R_j^*$ (as $j < \tilde{j}$).
	Note that if ad $\ad_{i^*}$ was not matched in $\ALG_j$ then $\Gamma_{i^*} = 0$,
	or otherwise, we increase the number of charges on $\Gamma_{i^*}$ by one. 
	
	\textbf{Case 2.2}. $\ALG_{j-1} = \ALG_{j}$. 
	The greedy choice and its inequalities from Case 2.1 still apply, but fail to produce a positive lower bound. 
	That is, 
	$g_{LB} = \rw_e - \Gamma_{i} - q R_j \le 0$ for each $e=(i,j)$.
	Therefore, it is sufficient to only pay $\Gamma_{i^*}$ for this case.~%
	
	In Case 2, we use each $\Gamma_i$ at most once because \OPT contains at most one edge incident to ad $\ad_i$, given the matching constraint.
	Furthermore, $\tau_i$ is non-decreasing after re-assigning 
	either ad $\ad_i$ (by design of \alggbackproxy),
	or other ads $\ad_{i'}$ (by \cref{lemma:R}),
	so the payments in prior iterations remain valid,
	completing the proof.
\end{proofE}

Note that the 2-approximation guarantee is tight for both \cref{alg:greedy-back} and \cref{alg:greedy-backproxy}, and 
this barrier exists also for the special case where $\q = 0$, that is, a \mwm instance.

\begin{propositionE}
	\cref{alg:greedy-back} and \cref{alg:greedy-backproxy} cannot do better than 2-approximation.
\end{propositionE}
\begin{proofE}
	Fix $\q=0$, and then \streamads is reduced to a maximum weighted matching problem (\mwm).
	It is well known that a greedy algorithm cannot do better than 2-approximation for \mwm.
	Concretely, let $\nV=2$.
	Create two ads $\ad_1, \ad_2$ with slots $\slots_1 = \{1,2\}$ and $\slots_2 = \{2\}$, respectively.
	Set rewards $\rw_{11} = \rw_{22} = 1$ and $\rw_{12} = 1+\epsilon$.
	Thus, a backwards-greedy algorithm yields a revenue of $1+\epsilon$ by assigning $\ad_1$ to the 2-nd slot,
	while the optimum assignment yields 2.
	The ratio approaches 2 for an arbitrary small $\epsilon$.
\end{proofE}

The time complexity for the \alggbackproxy algorithm is 
$\bigO(|\E| + \nV |\M|)$ 
where $|\M| = \min \{ \nV, \nads \}$.
The second term is due to the fact that we may need to compute $f_j(\M), j\in[\nV]$ if a re-assignment occurs.

\subsection{Natural greedy for \streamads}\label{sec:algs:streamads:oblivious}
\cref{alg:greedy-backproxy} uses a non-oblivious greedy criterion,
inspired by our novel decomposition in \cref{eq:decomp}. 
We now prove that %
\cref{alg:greedy-back} guided by the \emph{exact} marginal reward of an ad also results in a 2-approximation algorithm. %
This seemingly complicated case 
is
a direct consequence of our proof for \cref{alg:greedy-backproxy}.

\begin{corollaryE}\label{thm:greedy-streamads-oblivious}
	\cref{alg:greedy-back} returns a 2-approximation for the \streamads problem.
\end{corollaryE}
\begin{proofE}%
	The proof is similar to \cref{thm:greedy-streamads}, 
	except that we need a different inequality for the Case 2 therein.
	Though \cref{alg:greedy-back} does not use the values $\tau_i$, 
	we use such values here only for the analysis, and assume that \cref{alg:greedy-back} updates the values $\tau_i$ as from \cref{thm:greedy-streamads}.
	Recall that $\Gamma_i := \tau_i (1-\q)^{\sigma(i)-j}$.
	
	Suppose that at slot $j$, 
	$\OPT_{j-1} = \OPT_{j} + e^*$, where $e^* = (i^*, j)$.
	If our \ALG (re-)assigns edge $e=(i, j)$, then by the greedy criterion, %
	\begin{align*}
		g_{i} &\ge g_{i^*} \\
		\rw_{ij} -\q R_j - \kappa_{ij} &\ge \rw_{i^*j} -\q R_j - \kappa_{i^*j},
	\end{align*}
	where $g_i$ denotes the marginal reward of (re-)assigning ad $\ad_i$, and
	$\kappa_{ij} := \rw_{ij} -\q R_j - g_{i}$.
	By \cref{lemma:LB}, we have for any $i$,
	\[
		g_i \ge \rw_{ij} -\q R_j - \Gamma_i
		\quad\implies\quad
		\Gamma_i \ge \kappa_{ij}.
	\]
	Therefore, we can use both $\Gamma_{i^*}$ and $g_j$ to charge for $\rw_{i^*j}$. 
	That is,
	\begin{align*}
		g_j + \Gamma_{i^*} 
		&= \rw_{ij} -\q R_j - \kappa_{ij} + \Gamma_{i^*} \\
		&\ge \rw_{i^*j} -\q R_j - \kappa_{i^*j} + \Gamma_{i^*} \\
		&\ge \rw_{i^*j} -\q R_j^* - \kappa_{i^*j} + \Gamma_{i^*} \\
		&\ge \rw_{i^*j} -\q R_j^* = g^*_j,
	\end{align*}
	where the inequalities follow by 
	the greedy rule,
	the fact that $R_j < R_j^*$, and
	\cref{lemma:LB},
	respectively.
	
	The claim follows by 
	charging every $g^*_j$ to $g_j$ and $\{ \Gamma_i \}$, and
	noting that every $\Gamma_i$ is used at most once among all iterations.
	We omit the details for the other cases, as they follow from \cref{thm:greedy-streamads}.
\end{proofE}

\subsection{Other practical algorithms}\label{sec:algs:others}
In this section, we introduce 
various algorithms for the \streamads problem, including 
enhanced variants of existing algorithms (from \citep{ieong2014advertising}), and multiple practical heuristics.
We list all algorithms below, and discuss their important design choices. 

\smallskip
\noindent
\emph{Flow- and matching-based algorithms.}
\citet{ieong2014advertising} devised a 4-approximation algorithm \algflow by finding a maximum weighted matching with fixed weights.
That is, the matching only considers the decaying effects from items but not ads.
The key idea is to reduce the dynamic decaying effect of ad placement %
by limiting the number of allocated ads (i.e., the matching size) via an additional cardinality constraint.
In our evaluation, we implement the \algflow algorithm by a minimum-cost flow, 
as from its original paper.~%

We enhance the \algflow algorithm with greedy assignments over the slots not matched by the flow-based procedure,
such an algorithm is denoted by \algflowg.

We also introduce a natural heuristic \algmwm, mentioned in \cref{sec:case}.
\algmwm does not enforce a cardinality constraint to the matching size,
and is
implemented via a standard maximum-weighted matching algorithm.

\smallskip
\noindent
\emph{Global greedy algorithm.}
We introduce another natural algorithm \alggglobal that repeatedly allocates an ad to a slot that maximizes the marginal reward over all allocations, 
provided the reward being positive. %
This requires computing the marginal reward of every candidate allocation, with time complexity $\bigO(|\E|^2 |\M|)$, which is expensive.
We improve such computation by noting 
that the marginal reward of any possible allocation is non-increasing over time.
This 
can be used to perform
\emph{lazy evaluation} of the marginal reward, i.e., maintaining upper bounds to the actual rewards.
That is, we sort all candidate allocations by their rewards in a decreasing order using a heap, and
we complete a greedy step if the reward of the top allocation is greater than the %
upper bounds of all other candidate allocations.
Typically, only a few edge weights (i.e., marginal gains) need to be updated at each  greedy~iteration. %

\smallskip
\noindent
\emph{Online greedy algorithms.}
In \cref{sec:case}, we mention an online algorithm \alggforward that allocates an ad in real-time as a user browses its session.
Such an algorithm greedily assigns the most rewarding ad to the slot being processed.

In addition we also consider \alggonline, an online algorithm introduced by \citet{ieong2014advertising}. %
The idea is to pre-determine a threshold \Cthr, and 
for each slot, allocate the most rewarding ad if its reward is greater than \Cthr.
In our experiments, we validate some heuristics to determine the value of \Cthr, which is often difficult to obtain.

\section{Related work}\label{sec:related}

For lack of space we only discuss the most related work.
For discussion on further research that may be of interest
we refer the reader to \cref{app:related}.

\smallskip
\noindent
\emph{Native streaming advertising.}
The study of sequential ad allocations originates from simple cascade models~\citep{kempe2008cascade,aggarwal2008sponsored},
for which a dynamic-programming algorithm was developed.
However, when the reward of an ad depends on the slot position, more sophisticated algorithms are needed~\cite{ieong2014advertising}.
After the work by~\citet{ieong2014advertising}, several approaches have been proposed, discussed below.

\citet{gamzu2019advertisement} study a variant of native stream advertising, 
taking into account the distance between consecutive ads to avoid ad fatigue.
\citet{yan2020ads} present a practical solution with an industrial application, by maximizing the revenue while requiring that the total user engagement from organic items exceeds a given threshold.
\citet{liao2022cross} adopt a RL-based model to combine a list of content items and a list of ads to produce a user feed.
However, none of these works consider dynamic decay in attention caused by ads.

\smallskip
\noindent
\emph{Positive externalities in advertising.}
On a high level, the \streamads problem 
is based on a
form of negative externalities,
that is, the presence of an ad has a negative effect on future ads.
There has been extensive research on the opposite, i.e., positive externalities, in advertising.
One notable example is word-of-mouth marketing~\citep{kempe2003maximizing,hartline2008optimal},
where it is beneficial to offer products, even for free, to a small group of influencers at the beginning of an ad campaign, to attract more customers.

\smallskip
\noindent
\emph{Online matching.}
There is a rich body of work if externalities are not considered. %
For example, a standard model of position auctions such as~\citep{varian2007position} is based on the separability assumption, i.e., 
the probability an ad receives a click if placed in a position is simply the product of the quality scores associated to the ad and the position,
independent therefore of other ads.
Under such assumptions, the allocation problem can be treated as a matching problem, 
for which various algorithms have been developed. 
We refer the readers to some excellent surveys about matching for more details~\cite{mehta2007adwords,devanur2022online,huang2024online}.
Our greedy algorithms are partly inspired by a related streaming algorithm~\citep{feldman2009online};
however, as already mentioned, more sophisticated techniques are needed to handle externalities.

\section{Experimental evaluation}\label{sec:exp}
We provide the first comprehensive empirical study 
on the \streamads problem. 
We do not consider the \streamadsr problem, 
as it is a special case of the \streamads problem, and 
significantly less challenging given that it can be solved optimally by our~\alggback algorithm. %

\noindent
Our evaluation investigates the following key questions.
	
(1) How do the algorithms perform 
	by fixing the bipartite graph structure,
	and varying the weights of the rewards? (\cref{sec:exp:bip})	

(2) What is the impact of the problem parameters, such the quitting probability $\q$, the number of ads $\nads$, and slots $\nV$? (\cref{sec:exp:ablation})

(3) How do the algorithms perform for the task of native advertising in content feeds in two realistic scenarios? (\cref{sec:exp:ad})

Our source code is made public for reproducibility.{\code}

We now describe the datasets and baselines, while details on the environment 
are presented in \cref{app:exp}.
Note that, to enhance robustness, all results report averages over three independent runs.

\smallskip
\noindent
\emph{Datasets.}
To the best of our knowledge, 
high-quality public real datasets for native advertising are scarce, and existing work mostly uses proprietary data~\cite{yan2020ads,carrion2021blending,liao2022cross}.
Hence, we 
explored %
two distinct types of datasets for our evaluation.
The first type considers random weighted bipartite graphs.
Such data is very general, %
and provides a comprehensive benchmark for the various algorithms considered.
The second type of data is obtained by simulating a scenario of native advertising based on real anonymized ad data; more details are discussed in \cref{sec:exp:ad}.

\smallskip
\noindent
\emph{Algorithms and baselines.}
We evaluate the performance of our algorithms:
the proposed greedy algorithms \alggback (\cref{alg:greedy-back}) and \alggbackproxy (\cref{alg:greedy-backproxy}), and the practical global greedy algorithm \alggglobal.
Baseline algorithms consist of:
two online greedy algorithms \alggforward and \alggonline,
the flow-based algorithm {\algflow} and its augmented variant \algflowg, and
the matching-based algorithm \algmwm.
We set the threshold of \alggonline to be the best reward of an ad allocation to the first slot.
We refer the reader to \cref{sec:algs:others} for a detailed description of the above baselines.

\subsection{Experiments on synthesized bipartite graphs}\label{sec:exp:bip}

\begin{figure}[t]
	\centering
	\subcaptionbox{Symmetric random weighting \label{fig:exp:bip:sym}}[0.49\columnwidth]{
		{\includegraphics{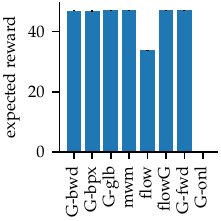}}
	}
	\hfill
	\subcaptionbox{Finely targeted weighting \label{fig:exp:bip:target}}[0.49\columnwidth]{
		{\includegraphics{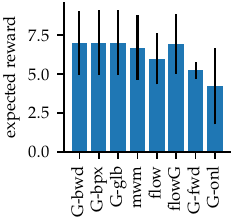}}
	}
	
	\subcaptionbox{Asymmetric random weighting (heavy tops) \label{fig:exp:bip:asym-top}}[0.49\columnwidth]{
		{\includegraphics{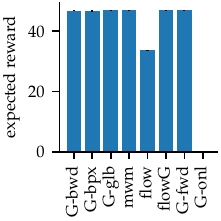}}
	}
	\hfill
	\subcaptionbox{Asymmetric random weighting (heavy bottoms) \label{fig:exp:bip:asym-bottom}}[0.49\columnwidth]{
		{\includegraphics{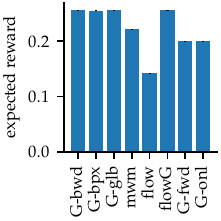}}
	}
	\caption{Comparisons on synthesized bipartite graphs with different weighting schemes. Error bars indicate the standard deviation.}
	\label{fig:exp:bip}
\end{figure}

In this setting, we first generate a fixed complete bipartite graph over
$\nads=100$ ads, and
$\nV=1000$ slots.
We evaluate the various algorithms
when the input instance has the following three different 
weighting schemes for the rewards over the edges of the graph:
1) symmetric random weighting,
2) asymmetric random weighting, and
3) finely targeted weighting.
Each setting is described in detail below.
We also fix $\q = 0.1$. %

\smallskip
\noindent
\emph{Symmetric random weighting.}
Each edge of the complete bipartite graph has its weight drawn uniformly at random from 1 to 10.

\smallskip
\noindent
\emph{Asymmetric random weighting.}
The random weighting scheme above
has symmetric edge weights for different slot positions,
which is not common in practice.
We break such symmetry and introduce dependencies with slot positions, by the following two methods.

In the first method, edges connecting a top slot have a larger reward.
More specifically, the reward $\rw_{ij}$ for assigning ad $\ad_i$ to slot $j$ is %
$\rw_{ij} = w \!\cdot\! (\nV - j) / \nV$, with $w$ a random real number in $[1,10]$, i.e., $\rw_{ij}$ likely decreases over increasing slot positions.
In the second method, edges connecting a bottom slot have a larger reward, that is, 
$\rw_{ij} = w \!\cdot\! j / \nV$.

\smallskip
\noindent
\emph{Finely targeted weighting.}
In practice, an ad may be highly relevant to just a few items.
To simulate this scenario,
for each ad $\ad_i$ we select a random slot $j$ and set $r_{ij}=10$, while setting $r_{ij'}=1$ for all other slots $j'\neq j$.

\smallskip
\noindent
\emph{Discussion.}
Results are reported in \cref{fig:exp:bip}.
We first note that the \alggonline\ algorithm has the worst performance, yielding zero reward on most instances. %
This is likely caused by the fact that its performances heavily depend on the threshold \Cthr, 
a parameter that is hard to optimize online.
In the current settings, a lower value  of \Cthr seems to lead to better solutions. %
The na\"{i}ve \alggforward algorithm, as expected, does not output good solutions if there are highly rewarding assignments for bottom slots.
In contrast, the \algmwm algorithm often outputs a solution with expected reward close to the best observed one, despite not accounting for decaying attention.
The 4-approximation algorithm \algflow achieves significantly lower expected rewards compared to the highest reward over all~algorithms. %

Our backwards greedy algorithms \alggback and \alggbackproxy,  \alggglobal, and \algflowg,
consistently outperform all other methods and achieve the highest expected reward over all settings,
with the global greedy algorithm \alggglobal providing slightly better solutions. %

\subsection{Ablation study}\label{sec:exp:ablation}

\begin{figure}[t]
	\centering
	
	\includegraphics{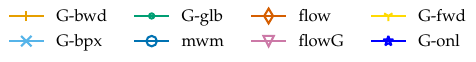}
	
	\subcaptionbox{Scaling the number of ads \label{fig:exp:ablation:ads}}[0.49\columnwidth]{
		{\includegraphics{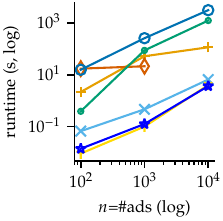}}
	}
	\hfill
	\subcaptionbox{Scaling the number of items \label{fig:exp:ablation:items}}[0.49\columnwidth]{
		{\includegraphics{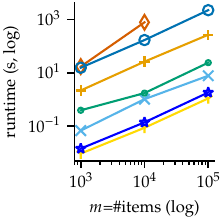}}
	}
	
	\subcaptionbox{Reward by varying \q \label{fig:exp:ablation:q-R}}[0.49\columnwidth]{
		{\includegraphics{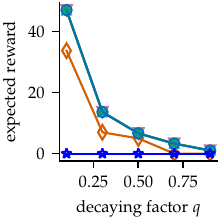}}
	}
	\hfill
	\subcaptionbox{Reward by varying $k$ \label{fig:exp:ablation:k-R}}[0.49\columnwidth]{
		{\includegraphics{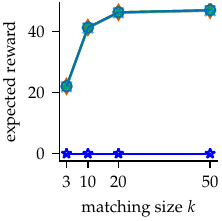}}
	}
	\caption{Effects of parameters $\nads, \nV, \q, k$.}
	\label{fig:exp:ablation}
\end{figure}

In this section, we investigate the effect of the various parameters, 
that may affect the performance of the algorithms.
We study 
the scalability with respect to the size of the bipartite graph, sensitivity to the decaying factor \q, and
to an additional cardinality constraint on the total number of ads to be displayed. %
We use the symmetric random weighting introduced previously for the edge weights.

\smallskip
\noindent
\emph{Scalability.}
We fixed $\q = 0.1$.
To test the scalability with respect to the input size, 
we start with
$\nads=100$ and
$\nV=1000$, and
vary the number of ads \nads and the number of videos \nV separately. %
The results are shown in \cref{fig:exp:ablation:ads} and \cref{fig:exp:ablation:items}, respectively.
We set a time limit of one hour for each run.
\algflow and \algmwm clearly 
have the largest running time, %
as they solve expensive optimization sub-problems.
Then, %
\alggglobal has also high running time, especially when $\nads$, the number of ads, grows, 
and is less sensitive to the number $\nV$ of slots due to the lazy evaluation of the rewards,
a technique 
we introduced in \cref{sec:algs:others}.
Considering our backwards-greedy algorithms \alggbackproxy and \alggback,
\alggbackproxy is significantly faster than \alggback,
since it uses a lower bound %
of the true marginal reward, achieving remarkable speedups. %
As expected, the two online algorithms are the fastest, at the expense of significantly lower rewarding solutions.

\smallskip
\noindent
\emph{Effect of \q.}
We fix the size of the complete bipartite graph, of ads and slots, to be
$\nads=100$ and 
$\nV=1000$, 
and we vary the parameter \q.
The result is shown in \cref{fig:exp:ablation:q-R}.
Clearly the expected reward drops as \q increases, as users are more likely to quit browsing early in the session.
We also note that the \algflow algorithm, cannot output a nonzero solution when $\q>0.5$; more details are on the original paper~\citep{ieong2014advertising}, making it not practical for general applications.

\smallskip
\noindent
\emph{Effect of size limit on ads.}
Given an integer $k$, we can adapt the algorithms to produce a matching of size \emph{at most} $k$ as follows.
We terminate the greedy \alggglobal and online algorithms after $k$  ad allocations.
We set the cardinality constraint of the \algflow algorithm to be exactly~$k$.
While, for all the other algorithms, we iteratively remove one ad at a time whose removal minimizes the loss in the expected reward, if more than $k$ slots are matched in their solution.

We fix $\nads=100$, $\nV=1000$ and $\q = 0.1$.
The result are in \cref{fig:exp:ablation:k-R}.
Overall, most algorithms obtain similar performance.
Moreover, as $k$ exceeds 20, their revenue reaches a plateau, and further ads bring unnoticeable benefit, in accordance with the value of $q$. %

\subsection{Simulated native advertising}\label{sec:exp:ad}

\begin{table}
	\caption{Datasets based on real advertisement. We report: $\nads$ number of ads to place, $\nV$ available slots, $|\E|$ number of edges, the range of the rewards and the value of $\q$ used in the experiments.}
	\label{tab:stats}
	\resizebox{0.99\columnwidth}{!}{
		\begin{tabular}{lcccrc}
			\toprule
			Dataset & $\nads$ & $\nV$ & $|\E|$ & $r_e$ ([$\min$ - $\max$]) & $\q$ \\
			\midrule
			YouTube & 120 & 14 999 & 1 799 880 & 2.9$\cdot 10^{-5}$ - 3.92$\cdot 10^5$& 0.1\\
			Criteo & 14 400 & 1 440 & 144 000 & 8.4$\cdot 10^0$ - 1.5$\cdot 10^3$& 0.1\\
			\bottomrule
		\end{tabular}
		}
\end{table}

\begin{figure}[t]
	\centering
	\subcaptionbox{Youtube dataset \label{fig:exp:real:yt}}[0.49\columnwidth]{
		{\includegraphics{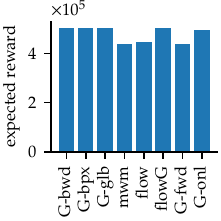}}
	}
	\hfill
	\subcaptionbox{Criteo dataset\label{fig:exp:real:criteo}}[0.49\columnwidth]{
		{\includegraphics{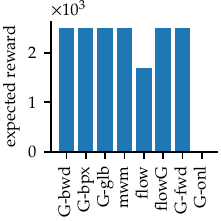}}
	}
	\caption{Comparisons on simulated native advertising using real data.}
	\label{fig:exp:real}
\end{figure}

\begin{figure}[t]
	\centering
	
	\includegraphics{pics/graph-legend-ncol4}
	
	\subcaptionbox{YouTube dataset \label{fig:exp:cumulYT}}[0.49\columnwidth]{
		{\includegraphics{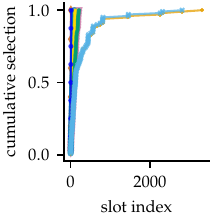}}
	}
	\hfill
	\subcaptionbox{Criteo dataset \label{fig:exp:cumulCrit}}[0.49\columnwidth]{
		{\includegraphics{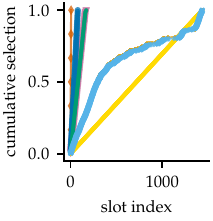}}
	}
	\caption{Distributions of the selected slot index. 
		}
	\label{fig:exp:cumulativeCurves}
\end{figure}

As mentioned previously, obtaining high-quality advertisement data is particularly challenging (given its proprietary nature). %
In this section we conduct experiments on two datasets built from real anonymous advertisement data, publicly available.

\smallskip
\noindent
\emph{Data generation.} 
Details on how we build instances to our problem based on two real-world datasets  
(videos from YouTube\footnote{\url{https://www.kaggle.com/datasets/sidharth178/youtube-adview-dataset}} and ads from the Criteo AI Lab\footnote{\url{https://go.criteo.net/criteo-research-kaggle-display-advertising-challenge-dataset.tar.gz}}) are in \cref{app:nativedata}.
Our instances successfully preserve the sequential and categorical distribution of advertisement rewards in the data, when available.
A summary of the key data statistics is reported in \cref{tab:stats}.

\smallskip
\noindent
\emph{Discussion.} 
First we report in \cref{fig:exp:real} the results, in terms of expected reward for the two datasets. 
We start by noting that on the YouTube dataset, the best performing algorithms %
are \alggback, \alggbackproxy, \alggglobal, and \algflowg, with \alggglobal outperforming all the other algorithms by a small margin. 
Surprisingly, the \alggonline algorithm also performs well.
Results for the Criteo dataset confirm a similar trend for the best performers, but this time together with \alggforward,
\alggonline performs poorly compared to others, given its very high sensitivity to \Cthr.
Such results are in line with what is observed on synthetic data, confirming the high quality solutions in output to our techniques. 

To further investigate the difference in the allocation strategies  produced by the algorithms, 
we analyzed how the various ads are placed over the slots. 
To do this, we report a cumulative distribution over the slot indices in output to each algorithm,
More specifically, suppose that an algorithm matches $k$ slots 
with indices $J \subseteq [\nV]$, then the cumulative value at index $j$ is $|\{j' \in J : j' \le j\}|/k$. 
The results are reported in \cref{fig:exp:cumulativeCurves}.
On the YouTube dataset, we observe very different allocation strategies.
We first note that methods with different ad allocation strategies may yield similar expected rewards, for example \alggglobal allocates more slots with larger indices than \algmwm despite achieving similar result on the Criteo dataset (see \cref{fig:exp:real:criteo}).
Our backwards greedy methods are the only ones that allocate ads to slots with large indices.
This is due to the backwards design,
which may allocate ads in bottom positions as long as they are beneficial,
even though their utility may diminish later.
In other words, our backwards greedy algorithms achieve a high recall rate of good allocations.
Ads with a diminished reward can be pruned with almost no loss in the final expected reward, e.g., by the pruning strategy we introduce in \cref{sec:exp:ablation}.

As a summary of our experiments, we observe that our proposed methods report high quality solutions with provable approximation guarantees (as captured by our analysis) on both synthetic and real data, and solve the \streamads problem much more efficiently than existing techniques.

\section{Conclusion}\label{sec:conclusion}
In this paper, we provide fast and practical 2-approximation greedy algorithms for the problem of advertising in content feeds.
We provide the first comprehensive empirical study on the problem, showing the strong performance of our methods. %
Regarding potential future work,
designing improved online algorithms, 
and studying alternative pricing schemes, dynamic rewards, and more flexible decaying functions 
are all interesting directions.

\balance

\begin{acks}
This research is supported by the
ERC Advanced Grant REBOUND (834862), 
the EC H2020 RIA project SoBigData++ (871042), and 
the Wallenberg AI, Autonomous Systems and Software Program (WASP) funded by the Knut and Alice Wallenberg Foundation.
\end{acks}

\bibliographystyle{ACM-Reference-Format}
\bibliography{references}


\begin{thebibliography}{31}


\ifx \showCODEN    \undefined \def \showCODEN     #1{\unskip}     \fi
\ifx \showISBNx    \undefined \def \showISBNx     #1{\unskip}     \fi
\ifx \showISBNxiii \undefined \def \showISBNxiii  #1{\unskip}     \fi
\ifx \showISSN     \undefined \def \showISSN      #1{\unskip}     \fi
\ifx \showLCCN     \undefined \def \showLCCN      #1{\unskip}     \fi
\ifx \shownote     \undefined \def \shownote      #1{#1}          \fi
\ifx \showarticletitle \undefined \def \showarticletitle #1{#1}   \fi
\ifx \showURL      \undefined \def \showURL       {\relax}        \fi
\providecommand\bibfield[2]{#2}
\providecommand\bibinfo[2]{#2}
\providecommand\natexlab[1]{#1}
\providecommand\showeprint[2][]{arXiv:#2}

\bibitem[Aggarwal et~al\mbox{.}(2008)]%
        {aggarwal2008sponsored}
\bibfield{author}{\bibinfo{person}{Gagan Aggarwal}, \bibinfo{person}{Jon
  Feldman}, \bibinfo{person}{Shanmugavelayutham Muthukrishnan}, {and}
  \bibinfo{person}{Martin P{\'a}l}.} \bibinfo{year}{2008}\natexlab{}.
\newblock \showarticletitle{Sponsored search auctions with markovian users}. In
  \bibinfo{booktitle}{\emph{International Workshop on Internet and Network
  Economics}}. Springer, \bibinfo{pages}{621--628}.
\newblock


\bibitem[Alhabash and Ma(2017)]%
        {Alhabash2017Socials}
\bibfield{author}{\bibinfo{person}{Saleem Alhabash} {and}
  \bibinfo{person}{Mengyan Ma}.} \bibinfo{year}{2017}\natexlab{}.
\newblock \showarticletitle{A Tale of Four Platforms: Motivations and Uses of
  Facebook, Twitter, Instagram, and Snapchat Among College Students?}
\newblock \bibinfo{journal}{\emph{Social Media + Society}} \bibinfo{volume}{3},
  \bibinfo{number}{1} (\bibinfo{date}{Jan.} \bibinfo{year}{2017}).
\newblock
\showISSN{2056-3051}
\href{https://doi.org/10.1177/2056305117691544}{doi:\nolinkurl{10.1177/2056305117691544}}


\bibitem[Buchbinder and Feldman(2018)]%
        {buchbinder2018submodular}
\bibfield{author}{\bibinfo{person}{Niv Buchbinder} {and} \bibinfo{person}{Moran
  Feldman}.} \bibinfo{year}{2018}\natexlab{}.
\newblock \showarticletitle{Submodular functions maximization problems}.
\newblock In \bibinfo{booktitle}{\emph{Handbook of approximation algorithms and
  metaheuristics}}. \bibinfo{publisher}{Chapman and Hall/CRC},
  \bibinfo{pages}{753--788}.
\newblock


\bibitem[Carrion et~al\mbox{.}(2021)]%
        {carrion2021blending}
\bibfield{author}{\bibinfo{person}{Carlos Carrion}, \bibinfo{person}{Zenan
  Wang}, \bibinfo{person}{Harikesh Nair}, \bibinfo{person}{Xianghong Luo},
  \bibinfo{person}{Yulin Lei}, \bibinfo{person}{Xiliang Lin},
  \bibinfo{person}{Wenlong Chen}, \bibinfo{person}{Qiyu Hu},
  \bibinfo{person}{Changping Peng}, \bibinfo{person}{Yongjun Bao},
  {et~al\mbox{.}}} \bibinfo{year}{2021}\natexlab{}.
\newblock \showarticletitle{Blending advertising with organic content in
  e-commerce: A virtual bids optimization approach}.
\newblock \bibinfo{journal}{\emph{arXiv preprint arXiv:2105.13556}}
  (\bibinfo{year}{2021}).
\newblock


\bibitem[Craswell et~al\mbox{.}(2008)]%
        {craswell2008experimental}
\bibfield{author}{\bibinfo{person}{Nick Craswell}, \bibinfo{person}{Onno
  Zoeter}, \bibinfo{person}{Michael Taylor}, {and} \bibinfo{person}{Bill
  Ramsey}.} \bibinfo{year}{2008}\natexlab{}.
\newblock \showarticletitle{An experimental comparison of click position-bias
  models}. In \bibinfo{booktitle}{\emph{Proceedings of the 2008 international
  conference on web search and data mining}}. \bibinfo{pages}{87--94}.
\newblock


\bibitem[Devanur and Mehta(2022)]%
        {devanur2022online}
\bibfield{author}{\bibinfo{person}{Nikhil Devanur} {and}
  \bibinfo{person}{Aranyak Mehta}.} \bibinfo{year}{2022}\natexlab{}.
\newblock \bibinfo{title}{Online matching in advertisement auctions}.
\newblock


\bibitem[eMarketer(2024)]%
        {stats4}
\bibfield{author}{\bibinfo{person}{eMarketer}.}
  \bibinfo{year}{2024}\natexlab{}.
\newblock \bibinfo{title}{US Native Advertising 2019}.
\newblock
\urldef\tempurl%
\url{https://www.emarketer.com/content/us-native-advertising-2019}
\showURL{%
\tempurl}
\newblock
\shownote{Accessed: Oct. 2024}.


\bibitem[Feldman et~al\mbox{.}(2009)]%
        {feldman2009online}
\bibfield{author}{\bibinfo{person}{Jon Feldman}, \bibinfo{person}{Nitish
  Korula}, \bibinfo{person}{Vahab Mirrokni},
  \bibinfo{person}{Shanmugavelayutham Muthukrishnan}, {and}
  \bibinfo{person}{Martin P{\'a}l}.} \bibinfo{year}{2009}\natexlab{}.
\newblock \showarticletitle{Online ad assignment with free disposal}. In
  \bibinfo{booktitle}{\emph{International workshop on internet and network
  economics}}. Springer, \bibinfo{pages}{374--385}.
\newblock


\bibitem[Gamzu and Koutsopoulos(2019)]%
        {gamzu2019advertisement}
\bibfield{author}{\bibinfo{person}{Iftah Gamzu} {and} \bibinfo{person}{Iordanis
  Koutsopoulos}.} \bibinfo{year}{2019}\natexlab{}.
\newblock \showarticletitle{Advertisement allocation and mechanism design in
  native stream advertising}. In \bibinfo{booktitle}{\emph{Complex Networks and
  Their Applications VII: Volume 2 Proceedings The 7th International Conference
  on Complex Networks and Their Applications COMPLEX NETWORKS 2018 7}}.
  Springer, \bibinfo{pages}{197--210}.
\newblock


\bibitem[Hartline et~al\mbox{.}(2008)]%
        {hartline2008optimal}
\bibfield{author}{\bibinfo{person}{Jason Hartline}, \bibinfo{person}{Vahab
  Mirrokni}, {and} \bibinfo{person}{Mukund Sundararajan}.}
  \bibinfo{year}{2008}\natexlab{}.
\newblock \showarticletitle{Optimal marketing strategies over social networks}.
  In \bibinfo{booktitle}{\emph{Proceedings of the 17th international conference
  on World Wide Web}}. \bibinfo{pages}{189--198}.
\newblock


\bibitem[Huang et~al\mbox{.}(2024)]%
        {huang2024online}
\bibfield{author}{\bibinfo{person}{Zhiyi Huang}, \bibinfo{person}{Zhihao~Gavin
  Tang}, {and} \bibinfo{person}{David Wajc}.} \bibinfo{year}{2024}\natexlab{}.
\newblock \showarticletitle{Online matching: A brief survey}.
\newblock \bibinfo{journal}{\emph{arXiv preprint arXiv:2407.05381}}
  (\bibinfo{year}{2024}).
\newblock


\bibitem[Ieong et~al\mbox{.}(2014)]%
        {ieong2014advertising}
\bibfield{author}{\bibinfo{person}{Samuel Ieong}, \bibinfo{person}{Mohammad
  Mahdian}, {and} \bibinfo{person}{Sergei Vassilvitskii}.}
  \bibinfo{year}{2014}\natexlab{}.
\newblock \showarticletitle{Advertising in a stream}. In
  \bibinfo{booktitle}{\emph{Proceedings of the 23rd international conference on
  World wide web}}. \bibinfo{pages}{29--38}.
\newblock


\bibitem[Kempe et~al\mbox{.}(2003)]%
        {kempe2003maximizing}
\bibfield{author}{\bibinfo{person}{David Kempe}, \bibinfo{person}{Jon
  Kleinberg}, {and} \bibinfo{person}{{\'E}va Tardos}.}
  \bibinfo{year}{2003}\natexlab{}.
\newblock \showarticletitle{Maximizing the spread of influence through a social
  network}. In \bibinfo{booktitle}{\emph{Proceedings of the ninth ACM SIGKDD
  international conference on Knowledge discovery and data mining}}.
  \bibinfo{pages}{137--146}.
\newblock


\bibitem[Kempe and Mahdian(2008)]%
        {kempe2008cascade}
\bibfield{author}{\bibinfo{person}{David Kempe} {and} \bibinfo{person}{Mohammad
  Mahdian}.} \bibinfo{year}{2008}\natexlab{}.
\newblock \showarticletitle{A cascade model for externalities in sponsored
  search}. In \bibinfo{booktitle}{\emph{International Workshop on Internet and
  Network Economics}}. Springer, \bibinfo{pages}{585--596}.
\newblock


\bibitem[Khanna et~al\mbox{.}(1998)]%
        {khanna1998syntactic}
\bibfield{author}{\bibinfo{person}{Sanjeev Khanna}, \bibinfo{person}{Rajeev
  Motwani}, \bibinfo{person}{Madhu Sudan}, {and} \bibinfo{person}{Umesh
  Vazirani}.} \bibinfo{year}{1998}\natexlab{}.
\newblock \showarticletitle{On syntactic versus computational views of
  approximability}.
\newblock \bibinfo{journal}{\emph{SIAM J. Comput.}} \bibinfo{volume}{28},
  \bibinfo{number}{1} (\bibinfo{year}{1998}), \bibinfo{pages}{164--191}.
\newblock


\bibitem[Kleinberg et~al\mbox{.}(2024)]%
        {kleinberg2024calibrated}
\bibfield{author}{\bibinfo{person}{Jon Kleinberg}, \bibinfo{person}{Emily Ryu},
  {and} \bibinfo{person}{{\'E}va Tardos}.} \bibinfo{year}{2024}\natexlab{}.
\newblock \showarticletitle{Calibrated recommendations for users with decaying
  attention}. In \bibinfo{booktitle}{\emph{International Symposium on
  Algorithmic Game Theory}}. Springer, \bibinfo{pages}{443--460}.
\newblock


\bibitem[Li et~al\mbox{.}(2024)]%
        {li2024deep}
\bibfield{author}{\bibinfo{person}{Xuejian Li}, \bibinfo{person}{Ze Wang},
  \bibinfo{person}{Bingqi Zhu}, \bibinfo{person}{Fei He},
  \bibinfo{person}{Yongkang Wang}, {and} \bibinfo{person}{Xingxing Wang}.}
  \bibinfo{year}{2024}\natexlab{}.
\newblock \showarticletitle{Deep automated mechanism design for integrating ad
  auction and allocation in feed}. In \bibinfo{booktitle}{\emph{Proceedings of
  the 47th International ACM SIGIR Conference on Research and Development in
  Information Retrieval}}. \bibinfo{pages}{1211--1220}.
\newblock


\bibitem[Liao et~al\mbox{.}(2022)]%
        {liao2022cross}
\bibfield{author}{\bibinfo{person}{Guogang Liao}, \bibinfo{person}{Ze Wang},
  \bibinfo{person}{Xiaoxu Wu}, \bibinfo{person}{Xiaowen Shi},
  \bibinfo{person}{Chuheng Zhang}, \bibinfo{person}{Yongkang Wang},
  \bibinfo{person}{Xingxing Wang}, {and} \bibinfo{person}{Dong Wang}.}
  \bibinfo{year}{2022}\natexlab{}.
\newblock \showarticletitle{Cross dqn: Cross deep q network for ads allocation
  in feed}. In \bibinfo{booktitle}{\emph{Proceedings of the ACM Web Conference
  2022}}. \bibinfo{pages}{401--409}.
\newblock


\bibitem[Meetanshi(2024)]%
        {stats3}
\bibfield{author}{\bibinfo{person}{Meetanshi}.}
  \bibinfo{year}{2024}\natexlab{}.
\newblock \bibinfo{title}{10 Native Advertising Statistics You Need to Know}.
\newblock
\urldef\tempurl%
\url{https://meetanshi.com/blog/native-advertising-statistics/}
\showURL{%
\tempurl}
\newblock
\shownote{Accessed: Oct. 2024}.


\bibitem[Mehta et~al\mbox{.}(2013)]%
        {mehta2013online}
\bibfield{author}{\bibinfo{person}{Aranyak Mehta} {et~al\mbox{.}}}
  \bibinfo{year}{2013}\natexlab{}.
\newblock \showarticletitle{Online matching and ad allocation}.
\newblock \bibinfo{journal}{\emph{Foundations and Trends{\textregistered} in
  Theoretical Computer Science}} \bibinfo{volume}{8}, \bibinfo{number}{4}
  (\bibinfo{year}{2013}), \bibinfo{pages}{265--368}.
\newblock


\bibitem[Mehta et~al\mbox{.}(2007)]%
        {mehta2007adwords}
\bibfield{author}{\bibinfo{person}{Aranyak Mehta}, \bibinfo{person}{Amin
  Saberi}, \bibinfo{person}{Umesh Vazirani}, {and} \bibinfo{person}{Vijay
  Vazirani}.} \bibinfo{year}{2007}\natexlab{}.
\newblock \showarticletitle{Adwords and generalized online matching}.
\newblock \bibinfo{journal}{\emph{Journal of the ACM (JACM)}}
  \bibinfo{volume}{54}, \bibinfo{number}{5} (\bibinfo{year}{2007}),
  \bibinfo{pages}{22--es}.
\newblock


\bibitem[Milano et~al\mbox{.}(2020)]%
        {Milano2020Recc}
\bibfield{author}{\bibinfo{person}{Silvia Milano},
  \bibinfo{person}{Mariarosaria Taddeo}, {and} \bibinfo{person}{Luciano
  Floridi}.} \bibinfo{year}{2020}\natexlab{}.
\newblock \showarticletitle{Recommender systems and their ethical challenges}.
\newblock \bibinfo{journal}{\emph{AI \& SOCIETY}} \bibinfo{volume}{35},
  \bibinfo{number}{4} (\bibinfo{date}{Feb.} \bibinfo{year}{2020}),
  \bibinfo{pages}{957--967}.
\newblock
\showISSN{1435-5655}
\href{https://doi.org/10.1007/s00146-020-00950-y}{doi:\nolinkurl{10.1007/s00146-020-00950-y}}


\bibitem[Outbrain(2022)]%
        {stats2}
\bibfield{author}{\bibinfo{person}{Outbrain}.} \bibinfo{year}{2022}\natexlab{}.
\newblock \bibinfo{title}{Top Native Advertising Statistics for 2022}.
\newblock
\urldef\tempurl%
\url{https://www.outbrain.com/blog/native-advertising-statistics}
\showURL{%
\tempurl}
\newblock
\shownote{Accessed: Oct. 2024}.


\bibitem[Shi et~al\mbox{.}(2023)]%
        {shi2023much}
\bibfield{author}{\bibinfo{person}{Yang Shi}, \bibinfo{person}{Jun~B Kim},
  {and} \bibinfo{person}{Ying Zhao}.} \bibinfo{year}{2023}\natexlab{}.
\newblock \showarticletitle{How much does ad sequence matter? Economic
  implications of consumer zapping and the zapping-induced externality in the
  television advertising market}.
\newblock \bibinfo{journal}{\emph{Journal of Advertising}}
  \bibinfo{volume}{52}, \bibinfo{number}{2} (\bibinfo{year}{2023}),
  \bibinfo{pages}{229--246}.
\newblock


\bibitem[Udwani(2023)]%
        {udwani2023submodular}
\bibfield{author}{\bibinfo{person}{Rajan Udwani}.}
  \bibinfo{year}{2023}\natexlab{}.
\newblock \showarticletitle{Submodular order functions and assortment
  optimization}. In \bibinfo{booktitle}{\emph{International Conference on
  Machine Learning}}. PMLR, \bibinfo{pages}{34584--34614}.
\newblock


\bibitem[Varian(2007)]%
        {varian2007position}
\bibfield{author}{\bibinfo{person}{Hal~R Varian}.}
  \bibinfo{year}{2007}\natexlab{}.
\newblock \showarticletitle{Position auctions}.
\newblock \bibinfo{journal}{\emph{international Journal of industrial
  Organization}} \bibinfo{volume}{25}, \bibinfo{number}{6}
  (\bibinfo{year}{2007}), \bibinfo{pages}{1163--1178}.
\newblock


\bibitem[Williamson and Shmoys(2011)]%
        {williamson2011design}
\bibfield{author}{\bibinfo{person}{David~P Williamson} {and}
  \bibinfo{person}{David~B Shmoys}.} \bibinfo{year}{2011}\natexlab{}.
\newblock \bibinfo{booktitle}{\emph{The design of approximation algorithms}}.
\newblock \bibinfo{publisher}{Cambridge university press}.
\newblock


\bibitem[Wojdynski and Golan(2016)]%
        {wojdynski2016native}
\bibfield{author}{\bibinfo{person}{Bartosz~W Wojdynski} {and}
  \bibinfo{person}{Guy~J Golan}.} \bibinfo{year}{2016}\natexlab{}.
\newblock \showarticletitle{Native advertising and the future of mass
  communication}.
\newblock \bibinfo{journal}{\emph{American Behavioral Scientist}}
  \bibinfo{volume}{60}, \bibinfo{number}{12} (\bibinfo{year}{2016}),
  \bibinfo{pages}{1403--1407}.
\newblock


\bibitem[Wu(2022)]%
        {wu2022submodular}
\bibfield{author}{\bibinfo{person}{Yizhan Wu}.}
  \bibinfo{year}{2022}\natexlab{}.
\newblock \bibinfo{title}{Submodular Order Maximization Subject to a p-Matchoid
  Constraint}.
\newblock


\bibitem[Yan et~al\mbox{.}(2020)]%
        {yan2020ads}
\bibfield{author}{\bibinfo{person}{Jinyun Yan}, \bibinfo{person}{Zhiyuan Xu},
  \bibinfo{person}{Birjodh Tiwana}, {and} \bibinfo{person}{Shaunak
  Chatterjee}.} \bibinfo{year}{2020}\natexlab{}.
\newblock \showarticletitle{Ads allocation in feed via constrained
  optimization}. In \bibinfo{booktitle}{\emph{Proceedings of the 26th ACM
  SIGKDD International Conference on Knowledge Discovery \& Data Mining}}.
  \bibinfo{pages}{3386--3394}.
\newblock


\bibitem[Yoon et~al\mbox{.}(2023)]%
        {yoon2023native}
\bibfield{author}{\bibinfo{person}{Hye~Jin Yoon}, \bibinfo{person}{Yan Huang},
  {and} \bibinfo{person}{Mark Yi-Cheon Yim}.} \bibinfo{year}{2023}\natexlab{}.
\newblock \showarticletitle{Native advertising relevance effects and the
  moderating role of attitudes toward social networking sites}.
\newblock \bibinfo{journal}{\emph{Journal of Research in Interactive
  Marketing}} \bibinfo{volume}{17}, \bibinfo{number}{2} (\bibinfo{year}{2023}),
  \bibinfo{pages}{215--231}.
\newblock


\end{thebibliography}

\ifsupp %
\clearpage
\appendix
\section{Missing proofs}\label{app:proofs}
\printProofs
\section{Native advertisement data}\label{app:nativedata}
In this section we describe how we built data used for our experimental evaluation on native advertisement, i.e., the setting in \cref{sec:exp:ad}.

\smallskip
\noindent
\emph{YouTube data}. The YouTube data we considered is formed by a set of videos $\{v_1, \dots, v_\nV\}$, characterized by: 
(1) the video category, i.e., $C(v_i) \in \{C_1,\dots,C_\ell\}$, where $\ell=8$; 
and (2) the number of ``ad views'' for each video, 
which we use as a proxy for the reward. %
To generate the data, we first obtain a random browsing session, i.e., 
a permutation $v_1',\dots,v_\nV'$ of the videos, through the following browsing model. 
A user starts from a randomly-chosen video $v_1'$. 
With probability $p=0.5$, the user selects another randomly chosen video of the same category $C(v_1')$, 
or otherwise the user randomly selects a previously unseen video from a different category.
The process is iterated until a permutation of all videos is obtained.

We assume that there are $r=15$ advertisers, providing $1,\dots,\ell$ ads, i.e., one for each category $k \in [\ell]$.  
We compute the reward $r_{ij}$ for ad $a_i$ after video $v_j$, where $i \in [r\ell]$ and $j \in[\nV]$, as follows. 
First, for each different category $C_k$ with $k \in [\ell]$, over all the videos belonging to $C_k$, 
we compute the average ``ad views'' $\mu_k$ and its standard deviation $\sigma_k$. 
We then assume that the rewards are normally distributed, i.e., 
$r_{ij} \sim \alpha_k |\mathcal{N}(\mu_k, \sigma_k)|$,
where $k = C(v_j)$, and
parameter $\alpha_k = 0.8$ if the ad and the video share the same category,
i.e., $C(\ad_i) = C(v_j)$,
or $\alpha_k=0.01$ otherwise, 
which captures a higher reward for ads targeted to related videos. 
Hence in the final data each ad $a_i, i\in [r\ell]$ can be placed after each video $v_j'$, 
with the reward $r_{ij}$ computed as above.

\smallskip
\noindent
\emph{Criteo data}. 
The data consists of a chronologically ordered \emph{sequence} of displayed ads collected over one day. 
Each of the 48~millions ads recorded has 13 numerical features (capturing the engagement of users with each displayed ad),
that we clustered into $k=100$ categories using the $k$-means algorithm. 
Besides, a reward can be computed for each ad, as a linear function of its features. %
We simulate the following browsing session over a full day: 
a user is browsing a website and an ad can be displayed to its session after one minute of content observed on the website, 
that is there are exactly $\nV=1440$ slots to which ads can be assigned.
We then create $b=144$ blocks of ads (which may correspond to different advertisers), and
for each block, we assume $k$ (non-existential) ads, i.e., one for each cluster.
We then associate ads in each block to 10 random slots among $\nV$.
Then, for each block-slot assignment we add connecting edges, that is, suppose the ads in block $h\in[b]$, with indices $a_{(h-1)k+1},\dots,a_{hk}$ are associated to slot $j$ then we add edges of the form $(a_{(h-1)k+i}, j)$ for $i\in[k]$. Then,
if there exists an edge between $\ad_i$ with $i \in [bk]$ and slot $j \in [\nV]$,
then the reward $r_{ij}$ %
is assumed to be the \emph{average reward}\footnote{More formally let $a_i=(a_i^1,\dots,a_i^{13})$ be ad $\ad_i$ with its features. Then we compute, for each ad it maximum engagement $\max_{h=1,\dots,13} \{|a_i^h|\}$, which we further multiply by a factor 10 if the ad was clicked by a user. Such value is then averaged to compute the actual average reward.} of all ads (from the original data) of the same category as $\ad_i$\footnote{among the $k$ categories obtained trough $k$-means.} displayed over the $j$-th minute; 
otherwise, $r_{ij} = 0$.
In this way, we capture the reward distribution over both clusters and time, in real-world data.

\section{Further related work}\label{app:related}
\smallskip
\noindent
\emph{Sequence submodularity.}
Although we show that the objective function of \streamads is non-monotone and non-submodular, 
it does obey a limited form of submodularity, that is, \prm-submodular order~\citep{udwani2023submodular,wu2022submodular}.
However, we cannot leverage such property without monotonicity. %
Moreover, the objective function also satisfies the so called ordered submodularity~\citep{kleinberg2024calibrated}.
Similarly, leveraging such stronger notion seems to be much harder.

\section{Experimental details}\label{app:exp}
\smallskip
\noindent
\emph{Environment.}
All algorithms are implemented in Python.
We adopt a solver for maximum flow and maximum matching from the NetworkX library.
All algorithms  are executed on a docker image of Ubuntu 22.04.
The server 
is hosted on a Linux system with  
48\,CPUs of Intel(R) Xeon(R) Gold 6336Y CPU @ 2.40\,GHz,
125\,GB RAM.%
\fi %

\end{document}